\documentclass[conference, a4paper]{IEEEtran}

\usepackage{comment}
\usepackage{cite}
\usepackage{graphicx}
\usepackage{epstopdf}
\usepackage{amsthm}
\usepackage[cmex10]{amsmath}
\usepackage{array}
\usepackage{algorithm}
\usepackage{algorithmic}
\usepackage{float}
\usepackage{graphicx}
\usepackage{amssymb}
\usepackage{microtype}
\usepackage{color}
\usepackage{relsize}
\usepackage{multirow}
\usepackage{balance}
\usepackage{amsmath}
\usepackage{hhline}
\usepackage[caption=false]{subfig}

\newtheorem{theorem}{Theorem}
\newtheorem{lemma}{Lemma}
\newtheorem{definition}{Definition}

\newtheorem{example}{Example}

\bstctlcite{IEEEexample:BSTcontrol}

\begin{document}

\title{A Finite-Length Construction of Irregular Spatially-Coupled Codes\vspace{-0.2cm}}

\author{\IEEEauthorblockN{Homa Esfahanizadeh, Ruiyi Wu, and Lara Dolecek}
\IEEEauthorblockA{Department of Electrical and Computer Engineering, University of California, Los Angeles, USA\\
hesfahanizadeh@ucla.edu, ruiyiwu@ucla.edu, and dolecek@ee.ucla.edu\vspace{-0.2cm}}}

\maketitle
%https://www.overleaf.com/6778473444smxnnpkzczbr
\begin{abstract}
Spatially-coupled (SC) LDPC codes have recently emerged as an excellent choice for error correction in modern data storage and communication systems due to their outstanding performance. It has long been known that irregular graph codes offer performance advantage over their regular counterparts. In this paper, we present a novel combinatorial framework for designing finite-length irregular SC LDPC codes. Our irregular SC codes have the desirable properties of regular SC codes thanks to their structure while offering significant performance benefits that come with the node degree irregularity. Coding constructions proposed in this work contribute to the existing portfolio of finite-length graph code designs.
\end{abstract}
\IEEEpeerreviewmaketitle

\section{Introduction}

Spatially-coupled (SC) LDPC codes are a class of high performance graph codes \cite{FelstromIT1999}. In the asymptotic limit of large code lengths, SC codes enjoy capacity achieving performance, and many elegant results have been derived in this setting, including threshold saturation, e.g., \cite{LentmaierIT2010,KudekarIT2013,AndriyanovaIT2016}.
In the finite-length regime, SC codes also offer excellent performance and are endowed with low-latency windowed decoding algorithms \cite{IyengarIT2013}. As a result, they have recently emerged as an excellent choice for error correction in modern data storage and communication systems  that require high levels of reliability \cite{EsfahanizadehTCOM2019,HareedyISIT2017}.

SC codes are constructed by coupling together disjoint (and identical) block codes into one long chain \cite{FelstromIT1999} as follows:  the parity-check matrix $\mathbf{H}$ of a block code is partitioned into component matrices $\mathbf{H}_k$, where $k\in\{0,1,\cdots,m\}$ and $m$ is the memory. Next, $L$ copies of component matrices, where $L$ is the coupling length, are pieced together to produce the parity-check matrix $\mathbf{H}_\text{SC}$ of the resultant SC code:\vspace{-0.3cm}

%\small
\begin{equation}
\label{SC_structure}
\mathbf{H}_\textnormal{SC}=\left[\begin{array}{cccccccc}
\mathbf{H}_0&\mathbf{0}&&&&&&\mathbf{0}\vspace{-0.2cm}\\
\mathbf{H}_1&\mathbf{H}_0&&&&&&\vdots\vspace{-0.2cm}\\
\vdots&\mathbf{H}_1&\ddots&&&&&\vspace{-0.2cm}\\
&\vdots&\ddots&&&&&\vspace{-0.2cm}\\
\mathbf{H}_m&\mathbf{H}_{m-1}&\ddots&&&&&\vspace{-0.2cm}\\
\mathbf{0}&\mathbf{H}_{m}&\ddots&&&&&\vspace{-0.2cm}\\
\vdots&\mathbf{0}&&&&&&\vspace{-0.2cm}\\
&\vdots&&&&&&\mathbf{H}_0\\
&&&&&&&\mathbf{H}_1\\
&&&&&&&\vdots\\
\mathbf{0}&\mathbf{0}&&&&&&\mathbf{H}_m\\
\end{array}\right]{.}
\end{equation}
%\normalsize

It has long been known that graph codes with irregular node degree distribution offer performance advantage over their regular-degree counterparts \cite{LubyIT2001}; this observation has led to the construction of many irregular \textit{block} LDPC codes with excellent properties, e.g., \cite{HuIT2005}, and design of capacity-approaching performance of highly irregular LDPC codes using density evolution techniques \cite{RichardsonIT2001}. While SC codes intrinsically possess a small amount of node degree irregularity due to the termination effects, which aides in performance improvement \cite{KudekarIT2013}, it is nonetheless customary to use regular underlying block to construct finite-length SC codes.

In this paper, we present a novel combinatorial framework for the finite length analysis and design of \textit{irregular} SC codes. We use circulant-based (CB) LDPC codes as the underlying block codes \cite{TannerIT2004}. We present a systematic scheme for optimal partitioning of the underlying block code and constructing irregular SC codes with a superior performance in the error-floor area. We show that this scheme has significantly better performance than random code constructions with the same node degree distribution. The performance advantage is achieved by explicitly organizing and combining circulants such that the population of graphical objects problematic for decoding is minimized.

%The rest of the paper is organized as follows. In Section~\ref{section:prelim}, we summarize the necessary background. In Section~\ref{section:const}, we show \textcolor{red}{ how irregular SC codes can be constructed by a new partitioning scheme that takes the irregularity into account}. In Section~\ref{section:part}, we present our new scheme for constructing irregular SC codes \textcolor{red}{through optimal partitioning}, which builds in part on an existing method for constructing regular SC codes, called optimal overlap partitioning (OO). Simulation results are presented in Section~\ref{section:sim}. Section~\ref{section:conc} delivers conclusions.\vspace{-0.0cm}

\section{Preliminaries}\label{section:prelim}
In this section, we review CB LDPC codes and SC codes derived by coupling CB LDPC codes. Then, we review the descriptions of certain graphical structures in the graph of LDPC codes which cause the error-floor phenomenon under iterative decoding, and we show that minimizing the population of instances  of the common subgraph that these problematic structures share in their configuration helps improve the error floor performance.

For a regular $(\gamma, \kappa)$ CB code, $\gamma$ is the column weight, i.e., variable node (VN) degree, and $\kappa$ is the row weight, i.e., check node (CN) degree. Let $z$ be the size of the constituent circulants. The parity-check matrix $\mathbf{H}$ of a CB code is as follows:
 
\begin{equation}
\mathbf{H}=\left[
\begin{array}{cccccccc}
\sigma^{f_{0,0}} & \sigma^{f_{0,1}}  & \dots & \sigma^{f_{0,\kappa-1}} \\
\sigma^{f_{1,0}} & \sigma^{f_{1,1}}  & \dots & \sigma^{f_{1,\kappa-1}} \\
\vdots & \vdots & \dots & \vdots\\
\sigma^{f_{\gamma-1,0}} & \sigma^{f_{\gamma-1,1}}  & \dots & \sigma^{f_{\gamma-1,\kappa-1}}
\end{array}
\right].
\end{equation}

Each VN corresponds to one column and each CN corresponds to one row in $\mathbf{H}$. Consider the parity-check matrix $\mathbf{H}$. We denote $i$, $0 \leq i \leq \gamma-1$, as the row group index and $j$, $0 \leq j \leq \kappa-1$, as the column group index. The circulant powers are non-negative integer values. If $f(i,j)=ij$ and $\kappa=z$, then this type of CB codes are also known as array-based (AB) codes \cite{FanTurbo2000}.%We use AB codes as underlying block codes in our simulation results.

The parity-check matrix $\mathbf{H}_\textnormal{SC}$ of an SC code has a band-diagonal structure. To construct a regular SC code, we partition the $\kappa \gamma$ circulants in the parity-check matrix $\mathbf{H}$ of a block CB code into component matrices $\mathbf{H}_k$, $0\leq k\leq m$. Each component matrix, which has the same size as $\mathbf{H}$, contains a subset of circulants from $\mathbf{H}$ and in the remaining entries has all zeros. Every circulant in $\mathbf{H}$ is assigned to exactly one of the component matrices, thus $\sum_{k=0}^{m}\mathbf{H}_k=\mathbf{H}$. Given the component matrices and parameter $L$, one can construct $\mathbf{H}_\textnormal{SC}$ as shown in (\ref{SC_structure}). Each group of columns of $\mathbf{H}_\textnormal{SC}$ that include one submatrix $[\mathbf{H}_0^T\dots\mathbf{H}_m^T]^T$, and zero elsewhere is called a replica. The $L$ replicas, from left to right, are denoted by $\mathbf{R}_1,\dots,\mathbf{R}_L$.

Certain structures in the graph of LDPC codes are responsible for most errors that occur under the iterative decoding in the error-floor region. Among the problematic graphical structures, the most harmful ones are absorbing sets (ASs) and trapping sets (TSs) \cite{DolecekIT2010,Richardson2003}. 
An $(a,b)$ TS is as a set of $a$ VNs in the graph of an LDPC code that are connected to exactly $b$ unsatisfied CNs if only the messages received by these $a$ VNs are erroneous. In case each VN is connected to strictly more satisfied than unsatisfied CNs, then the TS is an $(a,b)$ AS. ASs/TSs are  problematic objects in graph of irregular LDPC codes as well. Fig.~1 shows two examples of ASs in a graph of an irregular LDPC code. Based on our empirical results, we have identified that cycles with length $6$, i.e., cycles-$6$, are the common denominator instances of most problematic ASs/TSs for irregular SC codes with girth $6$, over AWGN channels.

In designing LDPC codes, we intend to minimize the population of ASs/TSs in order to have a better error-floor performance. Focusing on minimizing the population of instances of a small common subgraph, such as cycle-$6$, in the code design notably reduces the computational complexity and improves the performance. This observation follows from noting that the common denominator has a simpler graphical structure, and it exists in the graph of several detrimental ASs/TSs. 

\begin{figure}
\begin{tabular}{cc}
\includegraphics[width=0.19\textwidth]{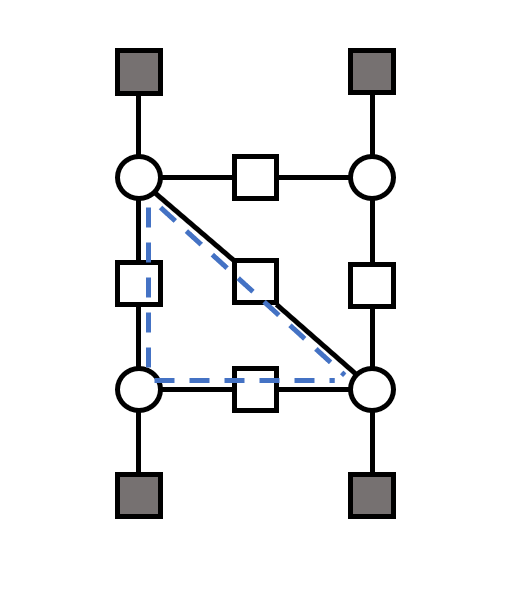}&
\includegraphics[width=0.19\textwidth]{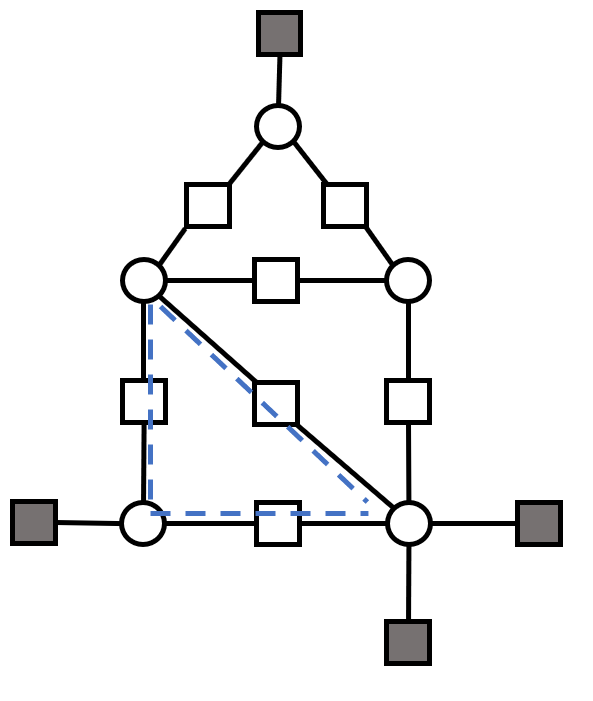}\\
(a)&(b)
\end{tabular}
\caption{Examples of ASs in graph of irregular LDPC codes along with their common denominator shown with dash blue line; (a) The $(4,4)$ AS. (b) The $(5,4)$ AS. Circles denote VNs, white squares denote satisfied CNs, and black squares denote unsatisfied CNs.\vspace{-0.5cm}}
\end{figure}

\section{Irregular SC Code Construction}\label{section:const}
As stated, regular SC codes are constructed by partitioning the circulants of a CB block code with non-zero circulants into $(m+1)$ component matrices. %For constructing irregular SC codes, one can choose an irregular underlying block code, i.e., a CB code with zero/non-zero circulants, for the partitioning. The resulting SC code has the same VN and CN degree distribution as the underlying block code, except for the first and last few of CNs.
In this section, we present a new procedure for constructing irregular SC codes that creates degree  irregularity in the partitioning step. For this purpose, we introduce a new component matrix,  which is called the \textit{dummy component matrix} and is denoted by  $\mathbf{H}_\textnormal{d}$. While $\mathbf{H}_\textnormal{d}$ is treated in the partitioning like the other component matrices,
\begin{equation}\label{equ:part}
    \mathbf{H}=\mathbf{H}_\textnormal{d}+\sum_{k=0}^{m}\mathbf{H}_k,
\end{equation}
the values assigned to $\mathbf{H}_\textnormal{d}$ are discarded, and $\mathbf{H}_\textnormal{d}$ does not appear in the parity-check matrix $\mathbf{H}_\textnormal{SC}$ of the irregular SC code, see (\ref{SC_structure}).

The protograph of a CB matrix is obtained by replacing each $z\times z$ non-zero circulant with scalar $1$ and each $z\times z$ zero circulant with scalar $0$, and it is denoted by the super-script $p$, i.e., $(.)^p$. For example, the protograph of $\mathbf{H}_k$, where $k\in\{0,\dots,m\}$, is $\mathbf{H}_k^p$. The degree distribution
%, i.e., the ratio of VNs (resp., CNs) with a specific degree value
is the same for a CB code and its protograph. Therefore, we define and evaluate the degree distribution for the protograph.
The VN (resp., CN) degree vector for the protograph of an irregular SC code is denoted by $\Lambda=[\lambda_0,\dots,\lambda_{\gamma-1}]$ (resp., $\Phi=[\rho_1,\dots,\rho_{\kappa-1}]$). Here, $\lambda_i$ (resp., $\rho_j$) is the portion of VNs (resp., CNs) with degree $i+1$ (resp., $j+1$). Parameter $\gamma$ (resp., $\kappa$) is the maximum VN (resp., CN) degree. The protograph of the dummy component matrix, i.e.,  $\mathbf{H}_\textnormal{d}^p$, determines the degree distribution of the final SC code, as shown by Lemma~\ref{degdist}.

\begin{lemma}\label{degdist}
Let $\alpha_{u}$, i.e., $0\leq u \leq \gamma-1$, be the number of $1$s in row $u$ of $\mathbf{H}_\textnormal{d}^p$. Similarly, let $\beta_{v}$, i.e., $0\leq v \leq \kappa-1$, be the number of $1$s in column $v$ of $\mathbf{H}_\textnormal{d}^p$. Then, $\forall i\in\{0,\dots,\gamma-1\}$ and $\forall j\in\{0,\dots,\kappa-1\}$,
\begin{equation}\label{VNCNdeg1}
\lambda_i=\frac{|\{\beta_{v}|0\leq v \leq\kappa-1,\beta_{v}=\gamma-i-1\}|}{\kappa},
\end{equation}
\begin{equation}\label{VNCNdeg2}
\rho_j\simeq\frac{|\{\alpha_{u}|0\leq u \leq \gamma-1,\alpha_{u}=\kappa-j-1\}|}{\gamma}.
\end{equation}
\end{lemma}
\begin{proof}
First, we derive the elements of the vector $\Lambda$. All replicas of $\mathbf{H}_\textnormal{SC}^p$ have the same non-zero parts. Thus, we only need to consider one replica, say $\mathbf{R}_1^p$, to derive the VN degree distribution. The $v$'th column in $\mathbf{R}_1^p$, where $0\leq v\leq\kappa-1$, has degree $\gamma-\beta_v$, see (\ref{SC_structure}) and (\ref{equ:part}). Therefore, the number of columns in $\mathbf{R}_1^p$ with degree $i+1$ is $|\{\beta_{v}|0\leq v \leq\kappa-1,\beta_{v}=\gamma-i-1\}|$. There are $\kappa$ columns in $\mathbf{R}_1^p$, and the expression in (\ref{VNCNdeg1}) follows. The elements of the vector $\Phi$, i.e.,  (\ref{VNCNdeg2}), can be derived similarly. The approximation sign in (\ref{VNCNdeg2}) is due to the fact that the first and the last group of CNs have lower degree due to the SC structure, (\ref{SC_structure}).\vspace{-0.1cm}
\end{proof}

\begin{example}
Assume the code parameters $\kappa=7$ and $\gamma=3$. We seek to construct $\mathbf{H}_\textnormal{d}^p$ to achieve  $\Lambda=[0,3/7,4/7]$ and $\Phi\simeq[0,0,0,0,0,1,0]$. Using (\ref{VNCNdeg1}) and (\ref{VNCNdeg2}), $\mathbf{H}_\textnormal{d}^p$ has $3$ columns with degree $1$, $4$ columns with degree =$0$, and $3$ rows with degree $1$. With these properties, one realization for $\mathbf{H}_\textnormal{d}^p$ is:
\begin{equation}
    \mathbf{H}_\textnormal{d}^p=\left[\begin{array}{ccccccc}
         1&0&0&0&0&0&0\\
         0&1&0&0&0&0&0\\ 
         0&0&1&0&0&0&0
    \end{array}\right].
\end{equation}
\end{example}
The node degree distributions, and consequently the matrix $\mathbf{H}_\textnormal{d}$, can be derived using the density evolution techniques to attain better threshold performance. In the next section, we present a new scheme for optimal partitioning of the circulant of $\mathbf{H}$ into the $(m+1)$ component matrices, considering the circulants that are already assigned to $\mathbf{H}_\textnormal{d}$, in order to minimize the number of cycles-$6$ in the protograph of an SC code.

\section{Optimal Partitioning for Irregular SC Codes}\label{section:part}

In this section, we extend a recently presented combinatorial approach for partitioning the underlying block code and constructing regular SC codes, called \textit{Optimal Overlap} (OO) partitioning \cite{EsfahanizadehTCOM2019}, for constructing irregular SC codes. For simplicity, we consider $m=1$ throughout this section. Thus, we intend to partition $\kappa\gamma$ circulants in matrix $\mathbf{H}$ of an underlying block code into $3$ component matrices $\mathbf{H}_{0}$, $\mathbf{H}_{1}$, and $\mathbf{H}_\textnormal{d}$, and piece $L$ copies of $\mathbf{H}_{0}$ and $\mathbf{H}_{1}$ in a diagonal structure to construct the irregular SC code. We note that the circulants that are assigned to $\mathbf{H}_\textnormal{d}$ are fixed given the degree distribution, and we aim to find the best partitioning of the remaining circulants between $\mathbf{H}_0$ and $\mathbf{H}_1$ such that the number of cycles-$6$ in the protograph of the SC code is minimized.

 For this purpose, we extend the definition of the overlap parameters, which are a set of integer-valued parameters that concisely include all necessary information needed to find the population of conbinatorial objects, e.g., cycles-$6$, in the protograph of an SC code. A careful selection of the overlap parameters corresponds to constructing high-performance irregular SC codes.

A cycle-$6$ in the graph of $\mathbf{H}_\textnormal{SC}^p$ results in either $z$ or $0$ cycles-$6$ in the graph of $\mathbf{H}_\textnormal{SC}$ depending on the powers of the circulants associated with that cycle \cite{Fossorier2004,Bazarsky2013}. Moreover, a cycle-$6$ in the final (lifted) graph of an SC code can only be generated from a cycle-$6$ in the protograph. Motivated by the above fact, our optimal partitioning aims at deriving the overlap parameters that result in the minimum number of the common denominator instances in the graph of $\mathbf{H}_\textnormal{SC}^p$.

We establish a discrete optimization problem by expressing the number of cycles-$6$ in the graph of $\mathbf{H}_\textnormal{SC}^p$ as a function of the overlap parameters and standard code parameters. We first review and extend the definition of overlap parameters for an irregular SC code:

\begin{definition}\label{def_ov_par}
Define matrix $\mathbf{\Pi}$ of size $3\gamma\times\kappa$ as follows:
\begin{equation}
\Pi=\left[\begin{array}{c}
\mathbf{H}_0^p\\
\mathbf{H}_1^p\\
\mathbf{H}_\textnormal{d}^p
\end{array}\right].
\end{equation}
A degree-$d$ overlap parameter $t_{\{i_1,\cdots,i_d\}}$, $0\leq i_1,\cdots,i_d \leq3\gamma-1$, is defined as the overlap among $d$ distinct rows of \text{ }$\mathbf{\Pi}$ specified by the set $\{i_1,\cdots,i_d\}$, i.e., the number of positions in which all the $d$ rows have $1$s simultaneously.
\end{definition}

Similar to the case of regular SC codes \cite{EsfahanizadehTCOM2019}, for an irregular SC code with maximum VN degree $\gamma$, the maximum degree for an overlap parameter with non-zero value is $\gamma$. Additionally, if there is at least one pair of distinct row indices $(i_u,i_v)$ such that $i_u,i_v\in\{i_1,\cdots,i_d\}$ and $i_u= i_v \text{ (mod $\gamma$)}$, then, $t_{\{i_1,\cdots,i_d\}}=0$.

\begin{lemma}
The set of non-deterministic independent overlap parameters $\mathcal{O}_\textnormal{ndi}$ is defined as follows:
\begin{equation}\label{equ_ind_nz_ov_par}
\begin{split}
\mathcal{O}_\textnormal{ndi}=&\{t_{\{i_1,\cdots,i_d\}}|1\leq d\leq \gamma,\gamma\leq i_1,\cdots,i_d\leq3\gamma-1,\\
&\forall \{i_u,i_v\}\subset\{i_1,\cdots,i_d\}\hspace{0.1cm}i_u\neq i_v \textnormal{ (mod $\gamma$)}\}\setminus\\
&\{t_{\{i_1,\cdots,i_d\}}|1\leq d\leq \gamma,2\gamma\leq i_1,\cdots,i_d\leq3\gamma-1\}.
\end{split}
\end{equation}
The other overlap parameters not in $\mathcal{O}_\textnormal{ndi}$ are either deterministic (zero or determined by the desired node degree distribution) or functions of the overlap parameters in $\mathcal{O}_\textnormal{ndi}$, as follows:

\begin{enumerate}
    \item Let $2\gamma\leq i_1,\cdots,i_{d} \leq3\gamma-1$. Then, $t_{\{i_1,\cdots,i_d\}}$ is determined based on the desired degree distribution.
    \item Let $0\leq i_1,\cdots,i_{d_1}\leq \gamma-1$, $\gamma \leq j_1,\cdots,j_{d_2}\leq 3\gamma-1$, and $1\leq(d_1+d_2)\leq \gamma$. Then, $t_{\{i_1,\cdots,i_{d_1},j_1,\cdots,j_{d_2}\}}$ is a linear function of the overlap parameters in $\mathcal{O}_\textnormal{ndi}:$
    \begin{equation}\label{equ_ind_nz_ov_par_relation}
\begin{split}
&t_{\{i_1,\dots,i_{d_1},j_1,\dots,j_{d_2}\}}=\\
&\hspace{0.5cm}t_\mathcal{J}+\sum\limits_{\alpha=1}^{d_1}{(-1)^\alpha}\hspace{-0.8cm}\sum\limits_{\substack{\{i_1',\dots, i_\alpha'\}\subset\mathcal{I}\\ [ x_1,\dots,x_\alpha ] \in\{1,2\}^\alpha}}\hspace{-0.7cm}t_{\mathcal{J}\cup\{x_1\gamma+i_1',\dots,x_{\alpha}\gamma+i_\alpha'\}},
\end{split}
\end{equation}
where $\mathcal{I}=\{i_1,\dots,i_{d_1}\}$, $\mathcal{J}=\{j_1,\dots,j_{d_2}\}$, and in the case of $\mathcal{J} = \varnothing$, $t_{\mathcal{J}} = \kappa$.
\end{enumerate}
\end{lemma}

\begin{proof} \text{ }
\begin{enumerate}
    \item Given the node degree distribution, the dummy component matrix $\mathbf{H}_\textnormal{d}$, its protograph, and its overlap parameters are determined.
    \item To have an overlap at position (column index) $y\in\{1,\dots\kappa\}$ among the rows $\mathcal{I}\cup\mathcal{J}$ of $\mathbf{\Pi}$: a) the rows in $\mathcal{J}$ of $\mathbf{\Pi}$ must have $1$s at position $y$, b) the rows in the second and third component matrices of $\mathbf{\Pi}$, i.e., $\mathbf{H}_1^p$ and $\mathbf{H}_\textnormal{d}^p$, that correspond to the rows in $\mathcal{I}$ must have $0$s at position $y$. In other words, the rows in $\{x_1\gamma+i_1,\dots,x_{d_1}\gamma+i_{d_1}\}$ of $\mathbf{\Pi}$ must have $0$'s at position $y$, where $[x_1,\dots,x_{d_1}]\in\{1,2\}^{d_1}$. Aided by the principle of inclusion and exclusion, (\ref{incexc}) follows.\vspace{-0.3cm}
\end{enumerate}
\end{proof}
\begin{example}\label{example_ind_ov_par}
For an irregular SC code with $m=1$, $\gamma=3$, $\kappa=7$, $\Lambda=[0,3/7,4/7]$, and $\Phi\simeq[0,0,0,0,0,1,0]$,
\begin{equation*}
\begin{split}
\mathcal{O}_\textnormal{ndi}&=\{t_3,t_4,t_5,t_{\{3,4\}},
t_{\{3,5\}},t_{\{4,5\}},t_{\{3,7\}},t_{\{3,8\}},t_{\{4,6\}},\\
&\hspace{0.65cm}t_{\{4,8\}},t_{\{5,6\}},t_{\{5,7\}},t_{\{3,4,5\}},t_{\{3,4,8\}},t_{\{3,5,7\}},\\
&\hspace{0.65cm}t_{\{3,7,8\}},t_{\{4,5,6\}},t_{\{4,6,8\}},t_{\{5,6,7\}}
\}. 
\end{split}
\end{equation*}
According to Lemma~2, the overlap parameters that are not in $\mathcal{O}_\textnormal{ndi}$ are deterministic or functions of the overlap parameters in $\mathcal{O}_\textnormal{ndi}$. For example,
\begin{equation*}
\begin{tabular}{ll}
\hspace{-0.2cm}$t_{\{6\}}=1$,&\hspace{-0.3cm}$t_{\{1\}}=6-t_{4}$,\\
\hspace{-0.2cm}$t_{\{78\}}=0$,&\hspace{-0.3cm}$t_{\{0,2\}}=5-t_{\{3\}}-t_{\{5\}}+t_{\{3,5\}}+t_{\{3,8\}}+t_{\{5,6\}}$,\\
\hspace{-0.2cm}$t_{\{678\}}=0$,&\hspace{-0.3cm}$t_{\{1,3\}}=t_{\{3\}}-t_{\{3,4\}}-t_{\{3,7\}}$.
\end{tabular}
\end{equation*}
\end{example}

%Lemma~\ref{lemma_num_in_nz_ov_par} describes the number of elements in $\mathcal{O}_\textnormal{ndi}$.
\begin{lemma}\label{lemma_num_in_nz_ov_par}
The number of non-deterministic independent overlap parameters is $\mathcal{N}_\textnormal{ndi}=|\mathcal{O}_\textnormal{ndi}|={3^\gamma-2^\gamma}$
\end{lemma}
\begin{proof}
Let the sets $T_1$ and $T_2$ be defined as follows\vspace{-0.05cm}:
\begin{equation*}
\begin{split}
T_1=&\{t_{\{i_1,\cdots,i_d\}}|1\leq d\leq \gamma,\gamma\leq i_1,\cdots,i_d\leq3\gamma-1,\\
&\forall \{i_u,i_v\}\subset\{i_1,\cdots,i_d\}\hspace{0.1cm}i_u\neq i_v \text{ (mod $\gamma$)}\},\\
T_2=&\{t_{\{i_1,\cdots,i_d\}}|1\leq d\leq \gamma,2\gamma\leq i_1,\cdots,i_d\leq3\gamma-1\}.
\end{split}
\end{equation*}
Since $\mathcal{O}_\textnormal{ind}=T_1\setminus T_2$, see (\ref{equ_ind_nz_ov_par}), and $T_2\subset T_1$, $\mathcal{N}_\textnormal{ind}=|\mathcal{O}_\textnormal{ind}|=|T_1|-|T_2|$. $|T_1|$ is the number of non-empty subsets of the set $S=\{\gamma,\dots,3\gamma-1\}$ with maximum size $\gamma$ such that no two elements in a subset have the same value mod $\gamma$. We partition the set $S$ into $\gamma$ disjoint sets $\{\gamma,2\gamma\},\{\gamma+1,2\gamma+1\},\dots,\{2\gamma-1,3\gamma-1\}$. The two elements in each of these partitions have the same value mod $\gamma$. Thus, we need to pick at most one element from each partition to form the set $T_1$, and there are $3^\gamma-1$ choices for this (selection of the first, second, or neither of elements for each partition; the case where none of the partitions lends an element to the subset, i.e., empty subset, must be excluded). $|T_2|$ is the number of non-empty subsets of the set $\{2\gamma,\dots,3\gamma-1\}$, i.e., $2^\gamma-1$. As a result,  $\mathcal{N}_\textnormal{ind}=3^\gamma-2^\gamma$.\vspace{-0.1cm}
\end{proof}

For example for $\gamma=3$, $\mathcal{N}_\textnormal{ndi}=19$. Building upon the results of \cite{EsfahanizadehTCOM2019}, we show that the number of cycles-$6$ in the protograph of an irregular SC code can  be expressed as a function of parameters in $\mathcal{O}_\textnormal{ndi}$.

\begin{theorem}\label{theorem_OO}
The number of cycles-$6$ in the protograph of an irregular SC code with parameters $m=1$, $L$, $\kappa$, and $\gamma$, and $\mathcal{O}_\textnormal{ndi}$ is:
\begin{equation}\label{equ_enumeration_proto}
F=LF_1^1+(L-1)F_1^2,
\end{equation}
and $F_1^1$ and $F_1^2$ are:
\begin{align}\label{equ_enumeration_proto_Fk}
F_1^1&=\hspace{-1cm}\sum_{\substack{
\{i_1,i_2,i_3\}\subset{\{0,\dots,2\gamma-1\}}\\
\overline{i_1}\neq\overline{i_2}\hspace{0.1cm},\overline{i_1}\neq\overline{i_3},\hspace{0.1cm}\overline{i_2}\neq\overline{i_3}}}\hspace{-1cm}
\mathcal{A}(t_{\{i_1,i_2,i_3\}},t_{\{i_1,i_2\}},t_{\{i_1,i_3\}},t_{\{i_2,i_3\}})\vspace{-0.15cm},\nonumber\\\text{}\nonumber \\
F_1^2&=\hspace{-0.85cm}\sum_{\substack{
i_1\in{\{0,\dots,2\gamma-1\}}\\
\{i_2,i_3\}\subset{\{\gamma,\dots,2\gamma-1\}}\\
\overline{i_1}\neq\overline{i_2}\hspace{0.1cm},\overline{i_1}\neq\overline{i_3},\hspace{0.1cm}\overline{i_2}\neq\overline{i_3}}}\hspace{-0.85cm}
\mathcal{B}(t_{\{i_1,i_2,i_3\}},t_{\{i_1,i_2\}},t_{\{i_1,i_3\}},t_{\{i_2-\gamma,i_3-\gamma\}})\nonumber \\
&+\hspace{-0.5cm}\sum_{\substack{
i_1\in{\{0,\dots,2\gamma-1\}}\\
\{i_2,i_3\}\subset{\{0,\dots,\gamma-1\}}\\
\overline{i_1}\neq\overline{i_2}\hspace{0.1cm},\overline{i_1}\neq\overline{i_3},\hspace{0.1cm}\overline{i_2}\neq\overline{i_3}}}\hspace{-0.6cm}
\mathcal{B}(t_{\{i_1,i_2,i_3\}},t_{\{i_1,i_2\}},t_{\{i_1,i_3\}},t_{\{i_2+\gamma,i_3+\gamma\}}),\nonumber
\end{align}
\normalsize
where $\overline{i}=(i\textnormal{ mod }\gamma)$. The functions $\mathcal{A}$ and $\mathcal{B}$ are defined in \cite{EsfahanizadehTCOM2019}, and are as follows:
\begin{equation*}
\begin{split}
\mathcal{A}&(t_{\{i_1,i_2,i_3\}},t_{\{i_1,i_2\}},t_{\{i_1,i_3\}},t_{\{i_2,i_3\}})\\
=&\left(t_{\{i_1,i_2,i_3\}}[t_{\{i_1,i_2,i_3\}}-1]^+[t_{\{i_2,i_3\}}-2]^+\right)\\
+&\left(t_{\{i_1,i_2,i_3\}}(t_{\{i_1,i_3\}}-t_{\{i_1,i_2,i_3\}})[t_{\{i_2,i_3\}}-1]^+\right)\\
+&\left((t_{\{i_1,i_2\}}-t_{\{i_1,i_2,i_3\}})t_{\{i_1,i_2,i_3\}}[t_{\{i_2,i_3\}}-1]^+\right)\\
+&\left((t_{\{i_1,i_2\}}-t_{\{i_1,i_2,i_3\}})(t_{\{i_1,i_3\}}-t_{\{i_1,i_2,i_3\}})t_{\{i_2,i_3\}}\right).\\
&\text{ }\\
\mathcal{B}&(t_{\{i_1,i_2,i_3\}},t_{\{i_1,i_2\}},t_{\{i_1,i_3\}},t_{\{i_2\pm\gamma,i_3\pm\gamma\}})\\
=&\left(t_{\{i_1,i_2,i_3\}}[t_{\{i_1,i_3\}}-1]^+t_{\{i_2\pm\gamma,i_3\pm\gamma\}}\right)\\
+&\left((t_{\{i_1,i_2\}}-t_{\{i_1,i_2,i_3\}})t_{\{i_1,i_3\}}t_{\{i_2\pm\gamma,i_3\pm\gamma\}}\right).
\end{split}
\end{equation*}
\end{theorem}
\begin{proof}
The proof has the same flow as the proof of Theorem~2 in \cite{EsfahanizadehTCOM2019} for the case $m=1$ because the construction shown in (\ref{SC_structure}) is the same for both regular and irregular SC codes. The set of overlap parameters needed to calculate $F$ in (\ref{equ_enumeration_proto}) is a subset of $\mathcal{O}_\textnormal{ndi}$. The overlap parameters that are related to the dummy component matrix $\mathbf{H}_\textnormal{d}^p$ do not directly play a role in finding $F$, and they only help to exploit the dependency between the overlap parameters for the protograph of $[\mathbf{H}_0^T,\mathbf{H}_1^T]^T$. Once the overlap parameters for the protograph of $[\mathbf{H}_0^T,\mathbf{H}_1^T]^T$ are determined, $F$ can be found using (\ref{equ_enumeration_proto}). 
\end{proof}

Let $F^*$ be the minimum number of cycles-$6$ in $\mathbf{H}_\textnormal{SC}^{\textnormal{p}}$. Thus, our discrete optimization problem is:
\begin{equation}\label{equ_F_opt}
F^*=\min_{\mathcal{O}_\textnormal{ndi}}F.
\end{equation}

Consider an underlying block code with parameters $\kappa$ and $\gamma$. In the partitioning, each circulant of the matrix of the underlying block code, i.e., $\mathbf{H}$, that is not assigned to the dummy component matrix $\mathbf{H}_\textnormal{d}$ can be assigned to $\mathbf{H}_0$ or $\mathbf{H}_1$, resulting in roughly $2^{\kappa\gamma}$ possible options. The goal is to choose a partitioning that results in the lowest number of cycles-$6$ in the protograph of an irregular SC code. We reduced the problem of finding the optimal partitioning for irregular SC codes into an optimization problem over $\mathcal{N}_\textnormal{ndi}=3^\gamma - 2^\gamma$ integer-valued overlap parameters.

As the second (optional) step of constructing irregular SC codes, we run a heuristic algorithm for adjusting the powers of non-zero circulants in one replica of $\mathbf{H}_\textnormal{SC}$. We remind that each change that is made to one replica is also applied to all other replicas to preserve the structure described in (\ref{SC_structure}). This algorithm is introduced in \cite{EsfahanizadehTCOM2019} for constructing regular SC codes and is called circulant
power optimizer (CPO).
In fact, an algebraic
condition must hold on the powers of a group of $6$ non-zero
circulants in order that they form a cycle-$6$ in a CB code \cite{Fossorier2004,Bazarsky2013}.
The CPO adjusts the powers of the problematic circulants, i.e., those that are involved in the most cycles-$6$, to
break the necessary condition for as many remaining cycles-$6$
as possible without creating
cycles-$4$.

\section{Simulation Results}\label{section:sim}

In this section, we compare the performance of our irregular SC codes with arbitrarily constructed irregular SC codes, where all codes have the same length, rate, and degree distribution. First, we describe the code parameters.
SC-Codes~1-3 are SC codes with parameters $\gamma=4$, $\kappa=z=13$, $m=1$, $L=10$, length $1{,}690$ bits, and design rate $0.66$.
All three codes are irregular with $\Lambda=[0,0,8/13,5/13]$ and $\Phi\simeq[0,\dots,0,1,0,0]$.

SC-Code~1 is constructed by the previous method of partitioning via a cutting vector, see \cite{MitchellISIT2014}, and AB circulant powers. SC-Code~2 is constructed by the optimal partitioning and AB circulant powers.
SC-Code~3 is constructed by the optimal partitioning and circulant powers obtained by the CPO algorithm.
The optimal partitioning is obtained by using the new systematic framework that we presented in this paper. Moreover, in order to reduce the computational complexity, the \textit{balanced} partitioning choices of circulants of $\mathbf{H}$, between $\mathbf{H}_0$ and $\mathbf{H}_1$, are considered in (\ref{equ_F_opt}), and the optimal ones are chosen to construct SC-Codes~2 and 3. The balanced partitioning also gives more freedom to the CPO to adjust the circulant powers and result in a better performance.

The partitioning matrix $\mathbf{PM}=[h_{i,j}]$ with dimension $\gamma\times\kappa$, is defined as follows: A circulant with row group index $i$ and column group index $j$ in $\mathbf{H}$ is assigned to $\mathbf{H}_\textnormal{d}$, resp., $\mathbf{H}_0$, and $\mathbf{H}_1$, if $h_{i,j}$ is $X$, resp., $0$, and $1$. The circulant power matrix $\mathbf{CM}=[f_{i,j}]$, with dimension $\gamma\times\kappa$, stores the powers of non-zero circulants. The partitioning matrices $\mathbf{PM}_1$ for SC-Code~1 and $\mathbf{PM}_{2,3}$ for SC-Code~2 and 3, and circulant power matrix $\mathbf{CM}_3$ for SC-Code~3 are given below:

\scriptsize
\begin{equation*}
\mathbf{PM}_1\hspace{-0.1cm}=\hspace{-0.1cm}\left[
\arraycolsep=3pt\def\arraystretch{1}
\begin{array}{ccccccccccccc}
X&0&0&1&X&1&1&1&1&1&1&1&1\\
0&X&0&0&0&X&1&1&1&1&1&1&1\\
0&0&X&0&0&0&X&0&1&1&1&1&1\\
0&0&0&X&0&0&0&X&0&0&0&1&1
\end{array}\right],
\end{equation*}
\scriptsize
\begin{equation*}
\mathbf{PM}_{2,3}\hspace{-0.1cm}=\hspace{-0.1cm}\left[
\arraycolsep=3pt\def\arraystretch{1}
\begin{array}{ccccccccccccc}
X&1&1&1&X&1&1&0&0&0&0&0&0\\
1&X&1&0&1&X&0&1&1&1&0&0&0\\
0&0&X&1&0&1&X&0&0&0&1&1&1\\
0&0&0&X&0&0&0&X&1&1&1&1&1
\end{array}\right],
\end{equation*}
\begin{equation*}
\mathbf{CM}_3\hspace{-0.1cm}=\hspace{-0.1cm}\left[
\arraycolsep=3pt\def\arraystretch{1}
\begin{array}{ccccccccccccc}
X&0&0&0&X&0&0&0&0&5&12&0&0\\
0&X&1&4&4&X&6&7&8&3&3&8&12\\
0&4&X&6&8&4&X&1&3&5&7&9&10\\
7&0&4&X&12&10&5&X&3&1&4&11&8
\end{array}\right].
\end{equation*}
\normalsize

In our simulations, we consider AWGN channels, and we use block min-sum algorithm with $15$ iterations for the decoding. Fig.~2 shows the FER performance for SC-Codes~1-3. As we see, SC-Code~3 shows $1.3$ and $0.7$ orders of magnitude performance improvement compared to SC-Codes~1 and 2 at SNR $=5.8\textnormal{ dB}$, respectively. In terms of the number of cycles, SC-Code~1, resp., SC-Code~2, and SC-Code~3, has $9{,}754$, resp., $4{,}397$, and $4{,}397$, cycles-$6$ in the protograph, and $12{,}896$, resp., $5{,}278$, and $1{,}469$, cycles-$6$ in the final graph.

\begin{figure}
\centering
\includegraphics[width=0.45\textwidth]{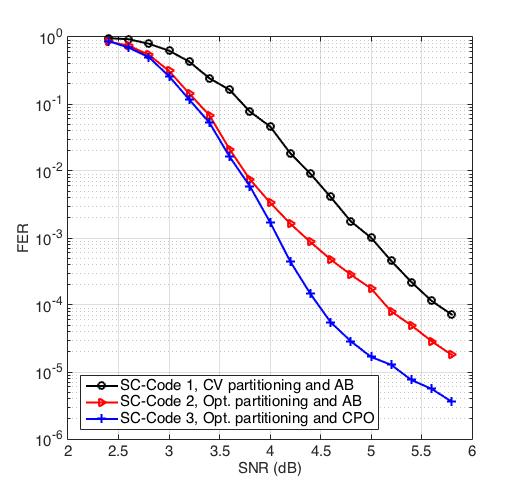}
\caption{FER comparison for irregular SC codes of the same length, rate, and degree distribution.\vspace{-1cm}}
\end{figure}

\section{Conclusion}\label{section:conc}
In this paper, we presented a novel combinatorial construction of finite-length irregular SC codes. Our code optimization is based on careful organization of circulants in the underlying block components. We showed on a representative example that our proposed codes offer performance improvement over comparable irregular SC codes designed used random assignments. Results from this work contribute to the growing portfolio of code constructions of finite-length SC codes. 

\section*{Acknowledgment}
Research supported in part by UCLA Dissertation Year Fellowship, Western Digital, and a grant CCF-BSF:CIF no. 1718389  from NSF.

\balance
\bibliographystyle{IEEEtran}
\bibliography{IrrSC_ITW}
%\bibliography{IEEEabrv,references}

\end{document}